\documentclass[3p,authoryear]{elsarticle}
\pdfoutput=1
\usepackage{tikz}
\usepackage[utf8]{inputenc}
\usepackage{amsmath}
\usepackage{nicefrac}
\usepackage[font=footnotesize,bf]{caption}
\newcommand{\final}{1}
\usepackage[colorlinks=true]{hyperref}
\usepackage{fancyhdr}
\usepackage{verbatim}
\usepackage{algorithm}
\usepackage{algpseudocode}
\usepackage{color}

\biboptions{square}

\newcommand{\codevar}[1]{\emph{#1}}

\newcommand{\D}[1]{}

\newcommand{\aand}{\textbf{and }}

\newcommand{\Break}{\textbf{\State exit loop}}
\newcommand{\Continue}{\textbf{repeat loop}}

\algrenewcommand\algorithmicindent{1.0em}
\algtext*{EndWhile} 
\algtext*{EndIf}    
\algtext*{EndFor}	  
\algtext*{EndProcedure}	
\newcommand{\LineIf}[2]{\State\algorithmicif\ {#1}\ \algorithmicthen\ {#2}}
\algrenewcommand\alglinenumber[1]{\footnotesize\normalfont{#1:}}

\makeatletter
\renewcommand{\ALG@beginalgorithmic}{\footnotesize}
\makeatother

\fancypagestyle{plain}{%
  \fancyhf{}
  \fancyhead[R]{\sc Drainage Over Flats}
  \cfoot{\thepage}
  
  }
\hypersetup{pdfauthor={Richard Barnes (ORCID: 0000-0002-0204-6040), Clarence Lehman, David Mulla},%
            pdftitle={An Efficient Assignment of Drainage Direction Over Flat Surfaces in Raster Digital Elevation Models},%
            pdfkeywords={flat resolution, terrain analysis, hydrology, drainage network, modeling, GIS},%
            pdfproducer={LaTeX},%
            pdfcreator={pdfLaTeX}
}

\definecolor{cg0}{HTML}{EDF8FB}
\definecolor{cg1}{HTML}{B2E2E2}
\definecolor{cg2}{HTML}{66C2A4}
\definecolor{cg3}{HTML}{238B45}

\definecolor{cr1}{HTML}{FFFFD4}
\definecolor{cr2}{HTML}{FED98E}
\definecolor{cr3}{HTML}{FE9929}
\definecolor{cr4}{HTML}{D95F0E}
\definecolor{cr5}{HTML}{993404}

\definecolor{cp1}{HTML}{FFF7F3}
\definecolor{cp2}{HTML}{FDE0DD}
\definecolor{cp3}{HTML}{FCC5C0}
\definecolor{cp4}{HTML}{FA9FB5}
\definecolor{cp5}{HTML}{F768A1}
\definecolor{cp6}{HTML}{DD3497}
\definecolor{cp7}{HTML}{AE017E}
\definecolor{cp8}{HTML}{7A0177}

\ifdefined\blackandwhite
  \tikzset{
    gcstyleaway0/.style = { fill=white },
    gcstyleaway1/.style = { fill=white },
    gcstyleaway2/.style = { fill=white },
    gcstyleaway3/.style = { fill=white },
    gcstyletowards1/.style = { fill=white },
    gcstyletowards2/.style = { fill=white },
    gcstyletowards3/.style = { fill=white },
    gcstyletowards4/.style = { fill=white },
    gcstyletowards5/.style = { fill=white },
    gcstylecombine1/.style = { fill=white },
    gcstylecombine2/.style = { fill=white },
    gcstylecombine3/.style = { fill=white },
    gcstylecombine4/.style = { fill=white },
    gcstylecombine5/.style = { fill=white },
    gcstylecombine6/.style = { fill=white },
    gcstylecombine7/.style = { fill=white },
    gcstylecombine8/.style = { fill=white }
  }
\else
  \tikzset{
    gcstyleaway0/.style = { fill=cg0 },
    gcstyleaway1/.style = { fill=cg1 },
    gcstyleaway2/.style = { fill=cg2 },
    gcstyleaway3/.style = { fill=cg3 },
    gcstyletowards1/.style = { fill=cr1 },
    gcstyletowards2/.style = { fill=cr2 },
    gcstyletowards3/.style = { fill=cr3 },
    gcstyletowards4/.style = { fill=cr4 },
    gcstyletowards5/.style = { fill=cr5 },
    gcstylecombine1/.style = { fill=cp1 },
    gcstylecombine2/.style = { fill=cp2 },
    gcstylecombine3/.style = { fill=cp3 },
    gcstylecombine4/.style = { fill=cp4 },
    gcstylecombine5/.style = { fill=cp5 },
    gcstylecombine6/.style = { fill=cp6 },
    gcstylecombine7/.style = { fill=cp7 },
    gcstylecombine8/.style = { fill=cp8 }
  }
\fi

\newcommand{\gcell}[5]{
	\fill [gcstyle#5#4] (0.5*#1,0.5*#2) rectangle (0.5*#1+0.5,0.5*#2+0.5);
  \node at (0.5*#1+0.25,0.5*#2+0.25) {\textbf{#3}};
}

\newcommand{\horizontallabels}[1]{
	\foreach \x in {1,...,7}{
		\node at (0.5*\x-0.25, #1) {\x};
	}
}

\newcommand{\verticallabels}[1]{
	\node at (#1,3.25) {A};
	\node at (#1,2.75) {B};
	\node at (#1,2.25) {C};
	\node at (#1,1.75) {D};
	\node at (#1,1.25) {E};
	\node at (#1,0.75) {F};
	\node at (#1,0.25) {G};
}

\newcommand{\gridcolor}{
	\fill [black!40] (0,0) rectangle (3.5,3.5);
	\fill [white] (0.5,0.5) rectangle (3,3);
	\fill [black!20] (1,0) rectangle (1.5,0.5);
}

\newcommand{\grid}{
	\draw[step=0.5cm,color=black] (0,0) grid (3.5,3.5);
}

\begin{document}
\thispagestyle{empty}

Cite as: Barnes, Lehman, Mulla. ``An Efficient Assignment of Drainage Direction Over Flat Surfaces In Raster Digital Elevation Models". Computers \& Geosciences. Vol 62, Jan 2014, pp 128--135. doi: ``10.1016/j.cageo.2013.01.009".

\begin{frontmatter}
  \title{An Efficient Assignment of Drainage Direction Over Flat Surfaces in Raster Digital Elevation Models}

	\author[rb]{Richard Barnes\corref{cor_rb}}
	\ead{rbarnes@umn.edu}
	\address[rb]{Ecology, Evolution, \& Behavior, University of Minnesota, USA}
	\cortext[cor_rb]{Corresponding author, 321-222-7637. ORCID: 0000-0002-0204-6040}

	\author[cl]{Clarence Lehman}
	\ead{lehman@umn.edu}
	\address[cl]{College of Biological Sciences, University of Minnesota, USA}

	\author[cl]{David Mulla}
	\ead{mulla003@umn.edu}
	\address[cl]{Soil, Water, and Climate, University of Minnesota, USA}

  \begin{abstract}
    \noindent In processing raster digital elevation models (DEMs) it is often necessary to assign drainage directions over flats---that is, over regions with no local elevation gradient. This paper presents an approach to drainage direction assignment which is not restricted by a flat's shape, number of outlets, or surrounding topography. Flow is modeled by superimposing a gradient away from higher terrain with a gradient towards lower terrain resulting in a drainage field exhibiting flow convergence, an improvement over methods which produce regions of parallel flow. This approach builds on previous work by Garbrecht and Martz (1997), but presents several important improvements. The improved algorithm guarantees that flats are only resolved if they have outlets. The algorithm does not require iterative application; a single pass is sufficient to resolve all flats. The algorithm presents a clear strategy for identifying flats and their boundaries. The algorithm is not susceptible to loss of floating-point precision. Furthermore, the algorithm is efficient, operating in $O(N)$ time whereas the older algorithm operates in $O(N^{\nicefrac{3}{2}})$ time. In testing, the improved algorithm ran 6.5 times faster than the old for a 100\,x\,100 cell flat and 69 times faster for a 700\,x\,700 cell flat. In tests on actual DEMs, the improved algorithm finished its processing 38--110 times sooner while running on a single processor than a parallel implementation of the old algorithm did while running on 16 processors. The improved algorithm is an optimal, accurate, easy-to-implement drop-in replacement for the original. Pseudocode is provided in the paper and working source code is provided in the Supplemental Materials.
  \end{abstract}

\end{frontmatter}


\section{Background}
A digital elevation model (DEM) is a representation of terrain elevations above some common base level, usually stored as a rectangular array of floating-point or integer values. With improvements in remote sensing, LIDAR, and computer performance, DEMs have increased in resolution from thirty-plus meters in the recent past to the sub-meter resolutions becoming available today. Increasing resolution has led to increased data sizes: current data sets are on the order of gigabytes and increasing, with billions of data points. While computer processing and memory performance have increased appreciably, legacy equipment and algorithms suited to manipulating smaller DEMs with coarser resolutions make processing improved data sources costly, if not impossible. Therefore, improved algorithms are needed. This paper presents one such algorithm.

DEMs may be used to estimate a region's hydrologic and geomorphic properties, including soil moisture, terrain stability, erosive potential, rainfall retention, and stream power. Many algorithms for extracting these properties require (1)~that every cell within a DEM have a defined flow direction and (2)~that by following flow directions from one cell to another, it is always possible to reach the edge of the DEM. These requirements are confounded by the presence of depressions and flats within the DEM.

Flats are mathematically level regions of the DEM with no local gradient. Although flats may occur naturally, their presence in a DEM is also frequently the result of technical issues in the DEM's collection and processing, such as from biased terrain reflectance, conversions from floating-point to integer precision, noise removal, low vertical resolution, or low horizontal resolution, among other possibilities.~\citep{Nardi2008,Garbrecht1997} Flats may also be produced when depressions---inwardly-draining regions of the DEM which have no outlet, also known as pits---are filled in. Depression-filling algorithms often increase the size and number of flats in a DEM, with one study finding 28--162\% more cells in flats after depressions were filled.~\citep{Nardi2008} In mountainous terrain the total number of cells in flats may be small, while in level or agricultural terrain the number may be a significant fraction of the DEM.

Ultimately, it is inevitable that the DEM will have flats. The algorithm by \citet{Garbrecht1997}---henceforth G\&M---described below presents one means of resolving them. G\&M is an improvement over previous algorithms (see \citet{Tribe1992}), offering a simple method to produce realistic flow patterns over flat terrain. More recent approaches using breadth-first search \citep{Liang2000}, priority-first search in weighted-graphs \citep{Jones2002}, cellular automata \citep{Coppola2007}, and variable elevation increments \citep{Jana2007}, among other methods (see \citet{Zhang2009}), may provide superior results but are often difficult to describe, implement, and test, leading to low adoption. \citet{Soille2003} is the only work of which the authors are aware which directly improves on G\&M (by eliminating the need for iterative applications of the algorithm). However, the method is described in terms of field-specific knowledge and little explanation of implementation is provided.

While most algorithms can be improved, G\&M's direct use in a number of studies (e.g.\ \citet{Alsdorf2003, Clarke2008, Clennon2007, Coe2008, Fang2010, Francesco2006, Kite2001, Lin2006, Phillips2007, White2004}), its inclusion in DEM processing packages such as TOPAZ \citep{Garbrecht1997} and TauDEM, as well as its inclusion as a preprocessing step in a number of other algorithms (e.g.\ \citet{Mackay1998, Miller2003, Stepinski2005, Tarboton2001, Toma2001, Turcotte2001, Vogt2003, Wallis2009, Zhang2009}), make it an important target for improvement.

This paper presents easy-to-implement improvements to G\&M which realize substantial gains in efficiency and accuracy.

\section{The Algorithm}
\subsection{Overview}
\label{sec:overview}
As noted by \citet{Garbrecht1997}, a flat will always be surrounded by two types of terrain: higher and lower. This is true even at the edges of the DEM because the improved algorithm assumes that the DEM's edge cells direct their flow outwards. It is natural to suppose that water near lower terrain will flow towards that lower terrain, whereas water near higher terrain will flow away from that higher terrain. This reasoning can be applied inductively to the entirety of a flat.

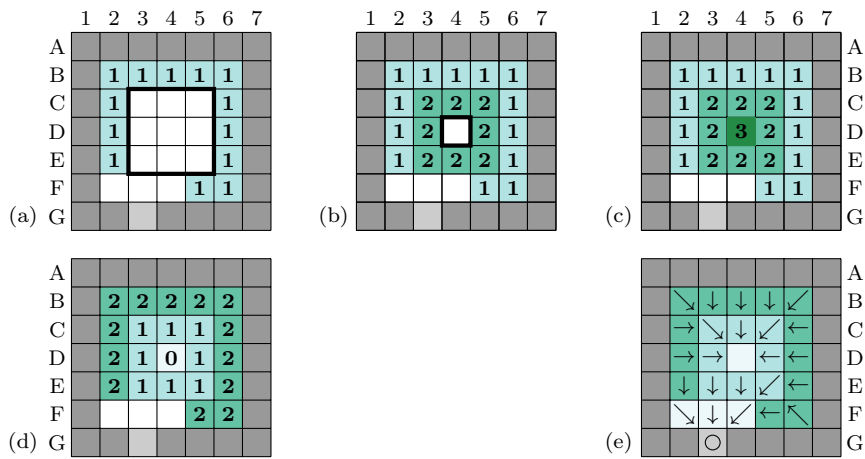
\begin{figure*}
	\centering
	\begin{tikzpicture}[scale=0.75]
	\tikzstyle{every node}=[font=\footnotesize]

	\newcommand{\panelone}{
		\gridcolor{}
    \gcell{4}{1}{1}{1}{away};
    \gcell{5}{1}{1}{1}{away};
		\foreach \x in {1,...,5}{
      \gcell{\x}{5}{1}{1}{away};
		}
		\foreach \x in {2,...,4}{
      \gcell{1}{\x}{1}{1}{away};
      \gcell{5}{\x}{1}{1}{away};
		}
    \grid{}
	}

	\newcommand{\paneltwo}{
    \gcell{2}{2}{2}{2}{away};
    \gcell{3}{2}{2}{2}{away};
    \gcell{4}{2}{2}{2}{away};

    \gcell{2}{4}{2}{2}{away};
    \gcell{3}{4}{2}{2}{away};
    \gcell{4}{4}{2}{2}{away};

    \gcell{2}{3}{2}{2}{away};
    \gcell{4}{3}{2}{2}{away};

    \grid{}
	}

	\newcommand{\panelthree}{
    \gcell{3}{3}{3}{3}{away};
	}

	\begin{scope}[shift={(0,0)}]
		\verticallabels{-0.25}
		\horizontallabels{3.75}
		\node at (-0.85, 0.25) {(a)};
		\panelone{}
		\draw[ultra thick] (1,1) rectangle (2.5,2.5);
	\end{scope}

	\begin{scope}[shift={(5.0,0)}]
		\node at (-0.4, 0.25) {(b)};
		\horizontallabels{3.75}
		\panelone{}
		\paneltwo{}
		\draw[ultra thick] (1.5,1.5) rectangle (2,2);
	\end{scope}

	\begin{scope}[shift={(10,0)}]
		\horizontallabels{3.75}
		\verticallabels{3.75}
		\node at (-0.4, 0.25) {(c)};
		\panelone{}
		\paneltwo{}
		\panelthree{}
	\end{scope}

	\begin{scope}[shift={(0,-4)}]
		\gridcolor{}
		\node at (-0.85, 0.25) {(d)};
		\verticallabels{-0.25}
    \gcell{4}{1}{2}{2}{away};
    \gcell{5}{1}{2}{2}{away};
		\foreach \x in {1,...,5}{
      \gcell{\x}{5}{2}{2}{away};
		}
		\foreach \y in {2,...,4}{
      \gcell{1}{\y}{2}{2}{away};
      \gcell{5}{\y}{2}{2}{away};
		}
    \gcell{2}{2}{1}{1}{away};
    \gcell{3}{2}{1}{1}{away};
    \gcell{4}{2}{1}{1}{away};

    \gcell{2}{4}{1}{1}{away};
    \gcell{3}{4}{1}{1}{away};
    \gcell{4}{4}{1}{1}{away};

    \gcell{2}{3}{1}{1}{away};
    \gcell{4}{3}{1}{1}{away};

    \gcell{3}{3}{0}{0}{away};
    \grid{}
	\end{scope}

	\begin{scope}[shift={(10,-4)}]
		\gridcolor{}
		\node at (-0.4, 0.25) {(e)};
		\verticallabels{3.75};

  \gcell{1}{5}{$\searrow$}{2}{away};
  \gcell{2}{5}{$\downarrow$}{2}{away};
  \gcell{3}{5}{$\downarrow$}{2}{away};
  \gcell{4}{5}{$\downarrow$}{2}{away};
  \gcell{5}{5}{$\swarrow$}  {2}{away};

  \gcell{1}{4}{$\rightarrow$}{2}{away};
  \gcell{2}{4}{$\searrow$}   {1}{away};
  \gcell{3}{4}{$\downarrow$} {1}{away};
  \gcell{4}{4}{$\swarrow$}   {1}{away};
  \gcell{5}{4}{$\leftarrow$} {2}{away};

  \gcell{1}{3}{$\rightarrow$}{2}{away};
  \gcell{2}{3}{$\rightarrow$}{1}{away};
  \gcell{3}{3}{}             {0}{away};
  \gcell{4}{3}{$\leftarrow$} {1}{away};
  \gcell{5}{3}{$\leftarrow$} {2}{away};

  \gcell{1}{2}{$\downarrow$}{2}{away};
  \gcell{2}{2}{$\downarrow$}{1}{away};
  \gcell{3}{2}{$\downarrow$}{1}{away};
  \gcell{4}{2}{$\swarrow$}  {1}{away};
  \gcell{5}{2}{$\leftarrow$}{2}{away};

  \gcell{1}{1}{$\searrow$}  {0}{away};
  \gcell{2}{1}{$\downarrow$}{0}{away};
  \gcell{3}{1}{$\swarrow$}  {0}{away};
  \gcell{4}{1}{$\leftarrow$}{2}{away};
  \gcell{5}{1}{$\nwarrow$}  {2}{away};
	\draw (1.25,0.25) circle (0.13);

  \grid{}

	\end{scope}
	\end{tikzpicture}
	\caption{Step~2: gradient away from higher terrain (see~\textsection\ref{sec:higher_grad} and Algorithm~\ref{alg:higher_grad}). The number of increments applied to each cell is shown. (a)~Numbers of increments after the first iteration. This first set of incremented cells are the edge cells adjacent to higher (but not to lower) terrain; they were found in Step~1 and stored in the queue \codevar{HighEdges}. After being incremented, each of these cells will add its unincremented neighbors---those just beyond the thick black line---to the queue. In (b) the neighbors of the edge cells have been popped off of the queue and incremented, advancing the black line. This process continues through (c), which shows the final number of increments, though the gradient is the inverse of what is desired. By noting that at most 3 increments were applied to this flat it is possible to obtain the desired gradient (d) by subtracting each cell of (c) from~3. This is done in Step~3 (Fig.~\ref{fig:combined_grad}). Note that the cells (F2, F3, F4) adjacent to lower terrain are ignored. (e)~shows the resulting drainage field. The flow directions of cells E3--E5 will be uniquely determined by a gradient towards lower terrain in Step~3 (see Fig.~\ref{fig:combined_grad}d).}
	\label{fig:higher_grad}
\end{figure*}

Using only the gradient away from higher terrain results in convergent, inward flow, as shown in Fig.~\ref{fig:higher_grad}e. Taken alone, this gradient would be unsuitable for further DEM processing because such flow does not in general drain from the flat. Using only the gradient towards lower terrain results in parallel flows which are guaranteed to drain the flat, as shown in Fig.~\ref{fig:lower_grad}f. Taken alone, this gradient would be unsuitable for further DEM processing because parallel flow patterns are uncommon in nature and, when present, tend to quickly evolve towards convergent flows. However, when these gradients combine, they produce a realistic, convergent flow pattern which is guaranteed to drain the flat.

If the flat is not adjacent to lower terrain, it cannot drain and therefore cannot be resolved by either the improved algorithm described in this paper or by G\&M. However, unlike G\&M, the improved algorithm will flag such flats and refuse to work on them. Aside from this requirement, the flat may be surrounded by any arbitrary combination of terrain.

Since the algorithm is used as a preprocessing step, its results must be available to subsequent processes. In G\&M this is accomplished by adding increments of $10^{-5}$ to the DEM's elevations. Ideally, these increments are negligably small compared to the vertical resolution of the DEM and are sufficient to direct flow while having no other impacts on the processing of the DEM. 

However, the precision of DEMs has increased considerably since G\&M was designed and it is now possible to find DEMs which use the full width of single-precision floating-point storage units. In such a case, there is no negligable increment which can be added---every increment is a significant alteration of the DEM. In this case, incrementing a cell to resolve a flat may cause it to rise above surrounding cells which were formerly higher than it was, corrupting important information about the landscape. Ultimately, G\&M provides no guarantees regarding its effect on a DEM.

Using double-precision floating-point storage is one solution, but, if a flat is particularly large or the DEM particularly precise, increments may still become significant. Using the larger data type also undesirably increases the storage size of the DEM in all subsequent operations.

Therefore, the improved algorithm generates an \textit{elevation mask} which is used to determine flow directions; the DEM's elevations themselves are left unaltered. If it is necessary to alter the DEM itself, a simple modification to the algorithm---discussed below---performs the alteration using what is guaranteed to be the smallest possible increment.

The algorithm assumes that the edge cells of the DEM which have defined elevations also have or can have defined flow directions---usually such flow is directed outwards, so the DEM drains itself. This assumption means that edge cells need not be treated as special cases.

The algorithm is divided into four steps, which are detailed below and described in pseudocode by Algorithm~\ref{alg:resolveflats}.
(1)~Each unique flat is identified and its edge cells grouped into two categories: those edge cells adjacent to higher terrain and those edge cells adjacent to lower terrain.
(2)~The gradient away from higher terrain will be constructed starting with the edge cells adjacent to higher terrain. During this step, the maximal number of increments for each flat will be noted.
(3)~The gradient towards lower terrain will be constructed starting with the edge cells adjacent to lower terrain. The results of Step~(2) are superimposed during this step.
(4)~The final flow directions are determined using the gradient developed in Steps~2 and~3.

\subsection{Step~1: Flat Identification}
\label{sec:flatid}
In order to guarantee that each flat drains and to calculate the gradient away from higher terrain in $O(N)$ time, the algorithm requires that each unique flat be identified.

The algorithm assumes that a separate flow direction algorithm of the user's choice has already been run and has assigned flow directions to each cell. Algorithm~\ref{alg:d8flowdirs} provides an example of how this might done. Some cells will have no place to drain to because they are surrounded by cells of the same or higher elevation; these cells will be given a special \textsc{NoFlow} value indicating that they do not have a defined flow direction. Let the data structure holding information regarding flow directions be called \codevar{FlowDirections}.

In order to seed the gradient routines, the algorithm requires a list of those cells of the flat which are adjacent to higher and lower terrain. To obtain this, the improved algorithm scans over each cell $c$ of \codevar{FlowDirections} (see Algorithm~\ref{alg:findedges}) and, where appropriate, adds $c$ to a queue for further processing.

$c$ is a ``high edge" cell if (1)~$c$ does not have a defined flow direction and (2)~$c$ has at least one neighbor $n$ at greater elevation. If these properties are true of $c$ it is added to the first-in, first-out (FIFO) queue \codevar{HighEdges}.

$c$ is a ``low edge" cell if (1)~$c$ has a defined flow direction, (2)~$c$ has at least one neighbor $n$ at the same elevation, and (3)~this $n$ does not have a defined flow direction. If these properties are true of $c$ it is added to the FIFO queue \codevar{LowEdges}.

Since every cell surrounding a flat must be either a low edge or a high edge cell, if none of either are found, then the DEM has no flats. If there are no low edge cells, but there are some high edge cells, then the DEM has flats but they cannot be drained. In both such instances, the algorithm indicates this and proceeds no further.

Otherwise, the algorithm then labels each unique flat (see Algorithm~\ref{alg:label}). Let the data structure holding these labels be called \codevar{Labels}, and let it be initialized to a value {\sc NoLabel}. To apply the labels, each cell $c$ in the \codevar{LowEdges} queue is used as the seed to a flood fill algorithm. If $c$ has not yet been labeled then a new label $l$ is given to it, its elevation $e$ is noted, and all cells which can be reached from $c$ while traversing only cells of elevation $e$ are given the same label $l$. A FIFO queue could be used for this, or a D8 variant of \citet{heckbert1990}, or one of many other solutions to the ``connected components" problem (\citet{grana2010} provide a good overview). Recursive solutions should be avoided in implementations using languages without proper tail recursion, as stack overflows are likely when processing large DEMs. In the end each flat will have a unique label and all member cells of a flat will bear its label.

The \codevar{LowEdges} queue will not contain any non-draining flats; however, such flats must be eliminated from the \codevar{HighEdges} queue. To do so, each cell in \codevar{HighEdges} is considered and, if its corresponding label is \codevar{NoLabel}, it is removed (this is correct since all high edge cells must eventually connect with a low edge cell and each low edge cell has been used as a labeling seed). If any cells are eliminated from \codevar{HighEdges}, it means that some of the DEM's flats will not drain.

If a flat does not drain, then it is a depression and may be resolved through depression-filling, among other methods. \emph{Barnes et al.\ }[in progress] describes an efficient algorithm for doing this in $O(m \log m)$ time, where $m\le N$; other less efficient solutions operating in $O(N \log N)$ time and higher are described by \citet{wang2006} and \citet{planchon2002}, respectively.

The results of Step~1 are (1)~two queues---\codevar{HighEdges} and \codevar{LowEdges}---containing all and only the drainable flat cells bordering regions of higher elevation and the flat cells bordering regions of lower elevation, respectively; and (2)~a structure \codevar{Labels} mapping each flat cell to a label indicating which flat it is a part of. At this point, all flats referred to by cells in either queue are guaranteed to drain following the completion of the algorithm.

	\begin{figure*}
		\centering
		\begin{tikzpicture}[scale=0.75]
		\tikzstyle{every node}=[font=\footnotesize]

		\newcommand{\panelone}{
			\gridcolor{}
      \gcell{1}{1}{1}{1}{towards};
      \gcell{2}{1}{1}{1}{towards};
      \gcell{3}{1}{1}{1}{towards};
		}

		\newcommand{\paneltwo}{
      \gcell{1}{2}{2}{2}{towards};
      \gcell{2}{2}{2}{2}{towards};
      \gcell{3}{2}{2}{2}{towards};
      \gcell{4}{2}{2}{2}{towards};
      \gcell{4}{1}{2}{2}{towards};
		}

		\newcommand{\panelthree}{
		  \foreach \x in {1,...,5}{
        \gcell{\x}{3}{3}{3}{towards};
		  }
      \gcell{5}{2}{3}{3}{towards};
      \gcell{5}{1}{3}{3}{towards};
		}

		\newcommand{\panelfour}{
		  \foreach \x in {1,...,5}{
        \gcell{\x}{4}{4}{4}{towards};
		  }
		}

		\newcommand{\panelfive}{
		  \foreach \x in {1,...,5}{
        \gcell{\x}{5}{5}{5}{towards};
		  }
		}

		\begin{scope}[shift={(0,0)}]
			\verticallabels{-0.25}
			\horizontallabels{3.75}
			\node at (-0.85, 0.25) {(a)};
			\panelone{}
      \grid{}
			\draw[ultra thick] (0.5,1) -- (2,1) -- (2,0.5);
		\end{scope}

		\begin{scope}[shift={(5.0,0)}]
			\node at (-0.4, 0.25) {(b)};
			\horizontallabels{3.75}
			\panelone{}
			\paneltwo{}
      \grid{}
			\draw[ultra thick] (0.5,1.5) -- (2.5,1.5) -- (2.5,0.5);
		\end{scope}

		\begin{scope}[shift={(10,0)}]
			\horizontallabels{3.75}
			\verticallabels{3.75}
			\node at (-0.4, 0.25) {(c)};
			\panelone{}
			\paneltwo{}
			\panelthree{}
      \grid{}
			\draw[ultra thick] (0.5,2) -- (3,2);
		\end{scope}

		\begin{scope}[shift={(0,-4)}]
			\verticallabels{-0.25}
			\node at (-0.85, 0.25) {(d)};
			\panelone{}
			\paneltwo{}
			\panelthree{}
			\panelfour{}
      \grid{}
			\draw[ultra thick] (0.5,2.5) -- (3,2.5);
		\end{scope}

		\begin{scope}[shift={(5,-4)}]
			\node at (-0.4, 0.25) {(e)};
			\panelone{}
			\paneltwo{}
			\panelthree{}
			\panelfour{}
			\panelfive{}
      \grid{}
		\end{scope}

		\begin{scope}[shift={(10,-4)}]
			\verticallabels{3.75}
			\gridcolor{}
			\node at (-0.4, 0.25) {(f)};

      \gcell{1}{5}{$\downarrow$}{5}{towards};
      \gcell{2}{5}{$\downarrow$}{5}{towards};
      \gcell{3}{5}{$\downarrow$}{5}{towards};
      \gcell{4}{5}{$\downarrow$}{5}{towards};
      \gcell{5}{5}{$\downarrow$}{5}{towards};

      \gcell{1}{4}{$\downarrow$}{4}{towards};
      \gcell{2}{4}{$\downarrow$}{4}{towards};
      \gcell{3}{4}{$\downarrow$}{4}{towards};
      \gcell{4}{4}{$\downarrow$}{4}{towards};
      \gcell{5}{4}{$\downarrow$}{4}{towards};

      \gcell{1}{3}{$\swarrow$}{3}{towards};
      \gcell{2}{3}{$\downarrow$}{3}{towards};
      \gcell{3}{3}{$\downarrow$}{3}{towards};
      \gcell{4}{3}{$\downarrow$}{3}{towards};
      \gcell{5}{3}{$\downarrow$}{3}{towards};

      \gcell{5}{2}{$\leftarrow$}{3}{towards};
      \gcell{4}{2}{$\swarrow$}{2}{towards};
      \gcell{3}{2}{$\downarrow$}{2}{towards};
      \gcell{2}{2}{$\downarrow$}{2}{towards};
      \gcell{1}{2}{$\downarrow$}{2}{towards};

      \gcell{5}{1}{$\leftarrow$}{3}{towards};
      \gcell{4}{1}{$\leftarrow$}{2}{towards};
      \gcell{3}{1}{$\swarrow$}{1}{towards};
      \gcell{2}{1}{$\downarrow$}{1}{towards};
      \gcell{1}{1}{$\searrow$}{1}{towards};

			\draw (1.25,0.25) circle (0.13);
      \grid{}

		\end{scope}
		\end{tikzpicture}
		\caption{Step~3: gradient towards lower terrain (see~\textsection\ref{sec:lower_grad} and Algorithm~\ref{alg:lower_grad}). The number of increments applied to each cell is shown. (a)~Numbers of increments after the first iteration. This first set of incremented cells are the edge cells adjacent to lower terrain; they were found in Step~1 and stored in the queue \textit{LowEdges}. After being incremented, each of these cells will add its unincremented neighbors---those just beyond the thick black line---to the queue. In (b) the neighbors of the edge cells have been popped off of the queue and incremented, advancing the black line. This process continues through (c), (d), and (e). (e)~shows the final number of increments. (f)~shows the resulting drainage field. There is a second part to Step~3, shown below, which combines the gradients.}
		\label{fig:lower_grad}
	\end{figure*}

	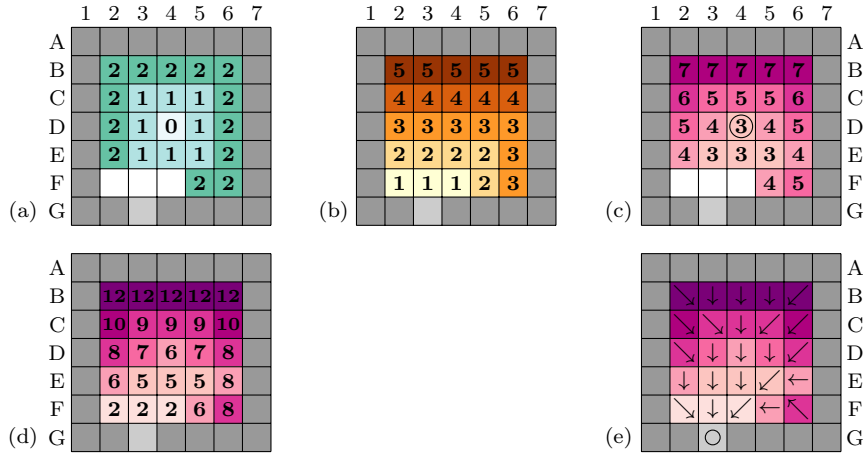
\begin{figure*}
		\centering
		\begin{tikzpicture}[scale=0.75]
		\tikzstyle{every node}=[font=\footnotesize]

		\newcommand{\panelaone}{
			\gridcolor{}
      \gcell{1}{1}{1}{1}{towards};
      \gcell{2}{1}{1}{1}{towards};
      \gcell{3}{1}{1}{1}{towards};
		}

		\newcommand{\panelatwo}{
		  \foreach \x in {1,...,4}{
        \gcell{\x}{2}{2}{2}{towards};
		  }
			\gcell{4}{1}{2}{2}{towards};
		}

		\newcommand{\panelathree}{
		  \foreach \x in {1,...,5}{
        \gcell{\x}{3}{3}{3}{towards};
		  }
      \gcell{5}{2}{3}{3}{towards};
      \gcell{5}{1}{3}{3}{towards};
		}

		\newcommand{\panelafour}{
		  \foreach \x in {1,...,5}{
        \gcell{\x}{4}{4}{4}{towards};
		  }
		}

		\newcommand{\panelafive}{
		  \foreach \x in {1,...,5}{
        \gcell{\x}{5}{5}{5}{towards};
		  }
		}

		\begin{scope}[shift={(5,0)}]
			\horizontallabels{3.75}
			\node at (-0.4, 0.25) {(b)};
			\panelaone{};
			\panelatwo{};
			\panelathree{};
			\panelafour{};
			\panelafive{};
      \grid{}
		\end{scope}

		\begin{scope}[shift={(0,0)}]
			\gridcolor{}
			\verticallabels{-0.25}
			\node at (-0.85, 0.25) {(a)};
			\horizontallabels{3.75}
			\gcell{4}{1}{2}{2}{away};
			\gcell{5}{1}{2}{2}{away};
			\foreach \x in {1,...,5}{
  			\gcell{\x}{5}{2}{2}{away};
			}
			\foreach \y in {2,...,4}{
  			\gcell{1}{\y}{2}{2}{away};
  			\gcell{5}{\y}{2}{2}{away};
			}

      \gcell{2}{2}{1}{1}{away};
      \gcell{3}{2}{1}{1}{away};
      \gcell{4}{2}{1}{1}{away};

      \gcell{2}{4}{1}{1}{away};
      \gcell{3}{4}{1}{1}{away};
      \gcell{4}{4}{1}{1}{away};

      \gcell{2}{3}{1}{1}{away};
      \gcell{4}{3}{1}{1}{away};

			\gcell{3}{3}{0}{0}{away};
      \grid{}
		\end{scope}

		\begin{scope}[shift={(10,0)}]
			\horizontallabels{3.75}
			\verticallabels{3.75}
			\gridcolor{}
			\node at (-0.4, 0.25) {(c)};

			\foreach \x in {1,...,5}{
  			\gcell{\x}{5}{7}{7}{combine};
			}

			\gcell{1}{4}{6}{6}{combine};
			\gcell{2}{4}{5}{5}{combine};
			\gcell{3}{4}{5}{5}{combine};
			\gcell{4}{4}{5}{5}{combine};
			\gcell{5}{4}{6}{6}{combine};

			\gcell{1}{3}{5}{5}{combine};
			\gcell{2}{3}{4}{4}{combine};
			\gcell{3}{3}{3}{3}{combine};
			\gcell{4}{3}{4}{4}{combine};
			\gcell{5}{3}{5}{5}{combine};
			\draw (1.75,1.75) circle (0.20);

			\gcell{1}{2}{4}{4}{combine};
			\gcell{2}{2}{3}{3}{combine};
			\gcell{3}{2}{3}{3}{combine};
			\gcell{4}{2}{3}{3}{combine};
			\gcell{5}{2}{4}{4}{combine};

			\gcell{4}{1}{4}{4}{combine};
			\gcell{5}{1}{5}{5}{combine};

      \grid{}
		\end{scope}

		\begin{scope}[shift={(0,-4)}]
			\node at (-0.85, 0.25) {(d)};
			\verticallabels{-0.25}
			\gridcolor{}

			\foreach \x in {1,...,5}{
  			\gcell{\x}{5}{\scriptsize 12}{8}{combine};
			}

			\gcell{1}{4}{\scriptsize 10}{7}{combine};
			\gcell{2}{4}{9}{6}{combine};
			\gcell{3}{4}{9}{6}{combine};
			\gcell{4}{4}{9}{6}{combine};
			\gcell{5}{4}{\scriptsize 10}{7}{combine};

			\gcell{1}{3}{8}{6}{combine};
			\gcell{2}{3}{7}{5}{combine};
			\gcell{3}{3}{6}{4}{combine};
			\gcell{4}{3}{7}{5}{combine};
			\gcell{5}{3}{8}{6}{combine};

			\gcell{1}{2}{6}{4}{combine};
			\gcell{2}{2}{5}{3}{combine};
			\gcell{3}{2}{5}{3}{combine};
			\gcell{4}{2}{5}{3}{combine};
			\gcell{5}{2}{8}{4}{combine};

			\gcell{1}{1}{2}{2}{combine};
			\gcell{2}{1}{2}{2}{combine};
			\gcell{3}{1}{2}{2}{combine};
			\gcell{4}{1}{6}{4}{combine};
			\gcell{5}{1}{8}{6}{combine};

      \grid{}
		\end{scope}

		\begin{scope}[shift={(10,-4)}]
			\verticallabels{3.75}
			\gridcolor{}
			\node at (-0.4, 0.25) {(e)};

      \gcell{1}{5}{$\searrow$}{8}{combine};
      \gcell{2}{5}{$\downarrow$}{8}{combine};
      \gcell{3}{5}{$\downarrow$}{8}{combine};
      \gcell{4}{5}{$\downarrow$}{8}{combine};
      \gcell{5}{5}{$\swarrow$}{8}{combine};

      \gcell{1}{4}{$\searrow$}{7}{combine};
      \gcell{2}{4}{$\searrow$}{6}{combine};
      \gcell{3}{4}{$\downarrow$}{6}{combine};
      \gcell{4}{4}{$\swarrow$}{6}{combine};
      \gcell{5}{4}{$\swarrow$}{7}{combine};

      \gcell{1}{3}{$\searrow$}{6}{combine};
      \gcell{2}{3}{$\downarrow$}{5}{combine};
      \gcell{3}{3}{$\downarrow$}{4}{combine};
      \gcell{4}{3}{$\downarrow$}{5}{combine};
      \gcell{5}{3}{$\swarrow$}{6}{combine};

      \gcell{1}{2}{$\downarrow$}{4}{combine};
      \gcell{2}{2}{$\downarrow$}{3}{combine};
      \gcell{3}{2}{$\downarrow$}{3}{combine};
      \gcell{4}{2}{$\swarrow$}{3}{combine};
      \gcell{5}{2}{$\leftarrow$}{4}{combine};

      \gcell{1}{1}{$\searrow$}{2}{combine};
      \gcell{2}{1}{$\downarrow$}{2}{combine};
      \gcell{3}{1}{$\swarrow$}{2}{combine};
      \gcell{4}{1}{$\leftarrow$}{4}{combine};
      \gcell{5}{1}{$\nwarrow$}{6}{combine};
      \grid{}
			\draw (1.25,0.25) circle (0.13);
		\end{scope}

		\end{tikzpicture}
		\caption{Step~3, continued: combining the gradients (see~\textsection\ref{sec:lower_grad} and Algorithm~\ref{alg:lower_grad}). (a)~The gradient away from higher terrain resulting from Step~2. (b)~The gradient towards lower terrain resulting from the first part of Step~3. (c)~Superposition of Steps 2 and Steps 3 in the style of G\&M, note that a new flat has been created at cell D4 (circled); this leads to flow direction ambiguity. (d)~Superposition as in the improved algorithm. By giving the gradient towards lower terrain twice the weight as the gradient away from higher terrain, all ambiguities are eliminated and convergent flow patterns produced. (e)~shows the resulting drainage pattern.}
		\label{fig:combined_grad}
	\end{figure*}

\subsection{Step~2: Gradient away from higher terrain}
\label{sec:higher_grad}
The gradient away from higher terrain is constructed by growing elevated terrain inwards from edge cells adjacent to higher terrain in a breadth-first expansion using a FIFO queue, a process described by Fig.~\ref{fig:higher_grad} and Algorithm~\ref{alg:higher_grad}.

To begin this process, a special ``iteration marker" is put at the end of the queue \codevar{HighEdges} developed in Step~1. An iteration counter $I$ is set to 1. Next, each cell $c$ is popped off of the queue in turn and its increment set to $I$ in a structure \codevar{FlatMask}. All of $c$'s unincremented neighbors which have the same label as $c$ (implying they are part of the same flat) and have undefined flow directions are added to the end of the queue. If $c$ is the iteration marker, then a new iteration marker is pushed onto the queue and $I$ is incremented by one.

Additionally, the maximal number of increments applied to each flat must be recorded. Therefore, an array \codevar{FlatHeights} with length equal to the number of flats is initialized to 0. When a cell is incremented, the cell's flat is looked up in \codevar{Labels} and the flat's corresponding value in \codevar{FlatHeights} is updated to $I$.

If a cell is adjacent to both lower and higher terrain, that cell is considered to be adjacent \emph{only} to the lower terrain and it is not considered in this step. The fulfillment of this constraint is guaranteed by Step~1.

Step~2 terminates when \codevar{HighEdges} contains only an iteration marker. The results of are (1)~a structure \codevar{FlatMask} indicating each cell's D8 distance away from higher ground and (2)~an array \codevar{FlatHeights} indicating the maximal distance any point in the flat is from higher ground. The values in \codevar{FlatMask} are the inverse of the desired gradient, but the values in \codevar{FlatHeights} can be used to invert the inverse, thereby producing the intended gradient (note the transition between Fig.~\ref{fig:higher_grad}c and Fig.~\ref{fig:higher_grad}d).

\subsection{Step~3: Gradient towards lower terrain}
\label{sec:lower_grad}
The gradient towards lower terrain is built via backgrowth from outlet(s) in the flat by working inwards from edge cells adjacent to lower terrain using a FIFO queue, a process described by Fig.~\ref{fig:lower_grad} and Algorithm~\ref{alg:lower_grad}.

This step is analogous to Step~2, with a few twists. First, every cell in \codevar{FlatMask} is made negative. Now, any cell which has a value of $0$ is not part of a flat and all cells which are negative have yet to be processed. Second, a special ``iteration marker" is put at the end of the queue \codevar{LowEdges} developed in Step~1. An iteration counter $I$ is set to 1.

As each cell $c$ is popped off of the queue, all of $c$'s unincremented neighbors which have the same label as $c$ and have undefined flow directions are added to the end of the queue. $c$ is then dealt with as specified below. If $c$ is the iteration marker, then a new iteration marker is pushed onto the queue and $I$ is incremented by one.

As each cell is popped, two things happen. If the cell's value in \textit{FlatMask} was incremented by the previous step, then it will have a negative value. The maximal height of the cell's flat---previously stored in \codevar{FlatHeights} in Step~2---is added to this value, thereby inverting the gradient away from higher terrain. Finally, the cell's value in \textit{FlatMask} is always incremented by $2I$. This is the twice the gradient towards lower terrain and enforces unambiguous drainage from the flat without the iterative applications required by G\&M, as discussed below.

Step~3 terminates when \codevar{LowEdges} contains only an iteration marker. The results are that \codevar{FlatMask} is the superposition of the gradient away from higher terrain and the gradient towards lower terrain.

\subsection{Step~4: Determination of Flow Directions}
\label{sec:flowdir}

In G\&M, exceptional situations occurred in which the resolution of flats created new flats (as in Fig.~\ref{fig:combined_grad}c); these situations were resolved by iterative application of G\&M using ever-smaller increments. However, loss of significance during this procedure could lead to unresolvable flats. In the improved algorithm presented in this paper, the gradient towards lower terrain is given twice the weight---made twice as steep---as the gradient away from higher terrain (see Fig.~\ref{fig:combined_grad}d). The gradient towards lower terrain, which is guaranteed to drain the flat, will then dominate ambiguous situations ensuring that all flats drain and that no iterative applications are necessary---a significant improvement over G\&M. \citet{Soille2003} propose a much more complex method of achieving the same end by using an ``inverse geodesic distance" as a mask, but provide few implementation details.

Finally, the flow directions algorithm is re-run on the DEM and flat cells with previously undefined flow directions are now defined using the increments stored in \codevar{FlatMask}. Because the increments are relative to the base elevation of a flat, cells from flats with differing labels must not be allowed to flow into each other unless their actual elevations permit it (which will only be the case for a flat's low edge cells). Flow directions can be determined using D8, D$\infty$ \citep{tarboton1997}, or any other such method.

Alternatively, if it is necessary to alter the DEM itself, as in G\&M, the smallest possible increments should be used. In \texttt{C} and \texttt{C++} this can be done using the \textsc{NextAfter} function defined by the \texttt{C99} and \texttt{POSIX} standards, which increases or decreases a floating-point number by the smallest possible increment in the direction of a second number; this is far preferable to an arbitrarily defined $\epsilon$. Similar functions exist in other languages. It is also possible to check that this process doesn't cause problems: after a cell is incremented, the only neighboring cells which should be lower than it are those which were lower before it was incremented, unless the neighbors in question are part of the same flat.

\section{Speed Comparisons}
Since the improved algorithm visits each cell only once per step, it therefore operates in $O(N)$ time. In contrast, it can be shown that G\&M operates in $O(N^2)$ time for long, narrow flats and in $O(N^{\nicefrac{3}{2}})$ time for square flats. Thus the improved algorithm described here can be orders of magnitude faster for large flats.

\begin{figure}
\centering
\footnotesize
\begin{tabular}{l c c c c l}
Algorithm & 100$^2$ cells & 400$^2$ cells & 700$^2$ cells & 1,000$^2$ cells \\
\hline
Improved	& 0.004\,s & 0.06\,s & 0.18\,s & 0.41\,s \\
G\&M	    & 0.026\,s & 1.82\,s & 12.5\,s & 45.4\,s \\
\hline
Speed-up  &   6.5\,x &   30\,x &   69\,x &  111\,x
\end{tabular}
\caption{Run-time comparisons of the improved algorithm versus the old. ``Speed-up" indicates how many times faster the improved algorithm ran than G\&M. Columns are labeled according to the number of cells in the test flat and run-times are in seconds. For comparison, in a 1m DEM an area of 100\,x\,100 cells is equivalent to 2.47~acres while an area of 1000\,x\,1000 cells is equivalent to 1\,km$^2$ or 100\,hectares. Tests were performed using \texttt{C++} implementations compiled with full optimizations using a 2.8GHz core. The implementations are included in the Supplemental Materials.}
\label{tbl:runtimes}
\end{figure}

The G\&M algorithm was tested against the improved algorithm on a DEM consisting of a large, square flat surrounded on all sides by higher elevation, except for a single-cell drainage in the bottom left, three cells in from the side---similar to the setup used in Fig.~\ref{fig:higher_grad}--\ref{fig:combined_grad}. Several test run times were averaged to yield the times shown in Fig.~\ref{tbl:runtimes} for DEMs of various sizes; standard deviations were small. The improved algorithm performed significantly faster than G\&M. The implementations used for this test are included in the Supplemental Materials.

\citet{Wallis2009} produced an MPI-based parallel implementation of G\&M for the TauDEM~5 analysis package. This implementation's flat resolution algorithm was tested against the improved algorithm included in the Supplemental Materials. Both implementations were built in \texttt{C++} and compiled with full optimizations; runs were performed on a two-node, 8-core, 2.8GHz machine.

In a test on a depression-filled 3\,m DEM of Minnesota's Beauford watershed (2418\,x\,1636 cells, 26\%~flats), the parallel implementation's fastest time was 20.37~seconds on 15~processors; the improved algorithm ran in 0.53~seconds on a single processor, 38~times faster in terms of elapsed time and 577~times faster in terms of processing time. On a depression-filled 3\,m DEM of Minnesota's Steele County (10891\,x\,13914 cells, 18\%~flats), the parallel implementation's fastest time was 53.25~minutes on 16~processors; the improved algorithm ran in 29~seconds, 110~times faster in terms of elapsed time and 1763~times faster in terms of total processing time. Both of the tested regions were agricultural landscapes consisting largely of fields.

\section{Coda}
An improved algorithm has been detailed to replace an existing algorithm by Garbrecht and Martz (1997). The improved algorithm produces the same result, but has a time complexity of only $O(N)$, whereas the old algorithm has a time complexity of $O(N^{\nicefrac{3}{2}})$ or worse; additionally, the improved algorithm performed significantly faster in actual situations. In addition, the improved algorithm does not require iterative application, is guaranteed to work only on flats which can be drained, and is not vulnerable to floating-point significance problems. Additionally, the improved algorithm is easier to implement than the old. Pseudocode is provided below. The Supplemental Materials contain a reference implementation, several DEMs, and the results of running the reference implementation on these DEMs for use in verifying custom implementations. The Supplemental Materials can be found in an archived format at \url{http://dx.doi.org/10.1016/j.cageo.2013.01.009} and in an interactive format at \url{https://github.com/r-barnes/Barnes2013-FlatSurfaces}.

This algorithm is implemented in the {\sc RichDEM} analysis package, 
which is available at \url{http://rbarnes.org/richdem} or via email from the authors.

\section{Acknowledgments}
Funding for this work was provided by the Minnesota Environment and Natural Resources Trust Fund under the recommendation and oversight of the Legislative-Citizen Commission on Minnesota Resources (LCCMR). Supercomputing time and data storage were provided by the Minnesota Supercomputing Institute. Joel Nelson provided LIDAR DEMs and ArcGIS support; Adam Clark provided feedback on the paper. In-kind support was provided by Justin Konen and Tess Gallagher. We are also grateful to the editors and to an anonymous reviewer.

{\footnotesize
	\bibliographystyle{elsarticle-harv.bst}
	\bibliography{refs}		

\begin{thebibliography}{32}
\expandafter\ifx\csname natexlab\endcsname\relax\def\natexlab#1{#1}\fi
\expandafter\ifx\csname url\endcsname\relax
  \def\url#1{\texttt{#1}}\fi
\expandafter\ifx\csname urlprefix\endcsname\relax\def\urlprefix{URL }\fi

\bibitem[{Alsdorf(2003)}]{Alsdorf2003}
Alsdorf, D., 2003. Water storage of the central amazon floodplain measured with
  gis and remote sensing imagery. Annals of the Association of American
  Geographers 93~(1), 55--66.
 doi: \url{http://dx.doi.org/10.1111/1467-8306.93105}

\bibitem[{Brardinoni and Hassan(2006)}]{Francesco2006}
Brardinoni, F., Hassan, M., 2006. Glacial erosion, evolution of river long
  profiles, and the organization of process domains in mountain drainage basins
  of coastal british columbia. Journal of geophysical research 111~(F1),
  F01013.
 doi: \url{http://dx.doi.org/10.1029/2005JF000358}

\bibitem[{Clarke et~al.(2008)Clarke, Burnett, and Miller}]{Clarke2008}
Clarke, S., Burnett, K., Miller, D., 2008. Modeling streams and hydrogeomorphic
  attributes in oregon from digital and field data. JAWRA Journal of the
  American Water Resources Association 44~(2), 459--477.
 doi: \url{http://dx.doi.org/10.1111/j.1752-1688.2008.00175.x}

\bibitem[{Clennon et~al.(2007)Clennon, King, Muchiri, and Kitron}]{Clennon2007}
Clennon, J., King, C., Muchiri, E., Kitron, U., 2007. Hydrological modelling of
  snail dispersal patterns in msambweni, kenya and potential resurgence of
  schistosoma haematobium transmission. Parasitology 134~(05), 683--693.
 doi: \url{http://dx.doi.org/10.1017/S0031182006001594}

\bibitem[{Coe et~al.(2008)Coe, Kinner, and Godt}]{Coe2008}
Coe, J., Kinner, D., Godt, J., 2008. Initiation conditions for debris flows
  generated by runoff at chalk cliffs, central colorado. Geomorphology
  96~(3-4), 270--297.
 doi: \url{http://dx.doi.org/10.1016/j.geomorph.2007.03.017}

\bibitem[{Coppola et~al.(2007)Coppola, Tomassetti, Mariotti, Verdecchia, and
  Visconti}]{Coppola2007}
Coppola, E., Tomassetti, B., Mariotti, L., Verdecchia, M., Visconti, G., 2007.
  Cellular automata algorithms for drainage network extraction and rainfall
  data assimilation. Hydrological sciences journal 52~(3), 579--592.
 doi: \url{http://dx.doi.org/10.1623/hysj.52.3.579}

\bibitem[{Fang et~al.(2010)Fang, Pomeroy, Westbrook, Guo, Minke, and
  Brown}]{Fang2010}
Fang, X., Pomeroy, J., Westbrook, C., Guo, X., Minke, A., Brown, T., 2010.
  Prediction of snowmelt derived streamflow in a wetland dominated prairie
  basin. Hydrology and Earth System Sciences 14~(6), 991--1006.
 doi: \url{http://dx.doi.org/10.5194/hessd-7-1103-2010}

\bibitem[{Garbrecht and Martz(1997)}]{Garbrecht1997}
Garbrecht, J., Martz, L., 1997. The assignment of drainage direction over flat
  surfaces in raster digital elevation models. Journal of hydrology 193~(1-4),
  204--213.
 doi: \url{http://dx.doi.org/10.1016/S0022-1694(96)03138-1}

\bibitem[{Grana et~al.(2010)Grana, Borghesani, and Cucchiara}]{grana2010}
Grana, C., Borghesani, D., Cucchiara, R., 2010. Optimized block-based connected
  components labeling with decision trees. IEEE Transactions on Image
  Processing 19~(6), 1596--1609.
 doi: \url{http://dx.doi.org/10.1109/TIP.2010.2044963}

\bibitem[{Heckbert(1990)}]{heckbert1990}
Heckbert, P., 1990. A seed fill algorithm. In: Graphics gems. Academic Press
  Professional, Inc., pp. 275--277.

\bibitem[{Jana et~al.(2007)Jana, Reshmidevi, Arun, and Eldho}]{Jana2007}
Jana, R., Reshmidevi, T., Arun, P., Eldho, T., 2007. An enhanced technique in
  construction of the discrete drainage network from low-resolution spatial
  database. Computers and geosciences 33~(6), 717--727.
 doi: \url{http://dx.doi.org/10.1016/j.cageo.2006.06.002}

\bibitem[{Jones(2002)}]{Jones2002}
Jones, R., 2002. Algorithms for using a dem for mapping catchment areas of
  stream sediment samples. Computers and geosciences 28~(9), 1051--1060.
 doi: \url{http://dx.doi.org/10.1016/S0098-3004(02)00022-5}

\bibitem[{Kite(2001)}]{Kite2001}
Kite, G., 2001. Modelling the mekong: hydrological simulation for environmental
  impact studies. Journal of Hydrology 253~(1-4), 1--13.
 doi: \url{http://dx.doi.org/10.1016/S0022-1694(01)00396-1}

\bibitem[{Liang and MaCkay(2000)}]{Liang2000}
Liang, C., MaCkay, D., 2000. A general model of watershed extraction and
  representation using globally optimal flow paths and up-slope contributing
  areas. International Journal of Geographical Information Science 14~(4),
  337--358.
 doi: \url{http://dx.doi.org/10.1080/13658810050024278}

\bibitem[{Lin et~al.(2006)Lin, Chou, Lin, Huang, and Tsai}]{Lin2006}
Lin, W., Chou, W., Lin, C., Huang, P., Tsai, J., 2006. Automated suitable
  drainage network extraction from digital elevation models in taiwan's
  upstream watersheds. Hydrological processes 20~(2), 289--306.
 doi: \url{http://dx.doi.org/10.1002/hyp.5911}

\bibitem[{Mackay and Band(1998)}]{Mackay1998}
Mackay, D., Band, L., 1998. Extraction and representation of nested catchment
  areas from digital elevation models in lake-dominated topography. Water
  resources research 34~(4), 897--901.
 doi: \url{http://dx.doi.org/10.1029/98WR00094}

\bibitem[{Miller(2003)}]{Miller2003}
Miller, D., 2003. Programs for dem analysis. Landscape dynamics and forest
  management. Gen. Tech. Rep. RMRSGTR-101CD. USDA Forest Service, Rocky
  Mountain Research Station, Fort Collins, Colorado, USA.[CD-ROM].

\bibitem[{Nardi et~al.(2008)Nardi, Grimaldi, Santini, Petroselli, and
  Ubertini}]{Nardi2008}
Nardi, F., Grimaldi, S., Santini, M., Petroselli, A., Ubertini, L., 2008.
  Hydrogeomorphic properties of simulated drainage patterns using digital
  elevation models: the flat area issue/propri{\'e}t{\'e}s
  hydro-g{\'e}omorphologiques de r{\'e}seaux de drainage simul{\'e}s {\`a}
  partir de mod{\`e}les num{\'e}riques de terrain: la question des zones
  planes. Hydrological Sciences Journal 53~(6), 1176--1193.
 doi: \url{http://dx.doi.org/10.1623/hysj.53.6.1176}

\bibitem[{Phillips and Slattery(2007)}]{Phillips2007}
Phillips, J., Slattery, M., 2007. Downstream trends in discharge, slope, and
  stream power in a lower coastal plain river. Journal of Hydrology 334~(1-2),
  290--303.
 doi: \url{http://dx.doi.org/10.1016/j.jhydrol.2006.10.018}

\bibitem[{Planchon and Darboux(2002)}]{planchon2002}
Planchon, O., Darboux, F., 2002. A fast, simple and versatile algorithm to fill
  the depressions of digital elevation models. Catena 46~(2-3), 159--176.
 doi: \url{http://dx.doi.org/10.1016/S0341-8162(01)00164-3}

\bibitem[{Soille et~al.(2003)Soille, Vogt, and Colombo}]{Soille2003}
Soille, P., Vogt, J., Colombo, R., 2003. Carving and adaptive drainage
  enforcement of grid digital elevation models. Water Resources Research
  39~(12), 1366--1375.
 doi: \url{http://dx.doi.org/10.1029/2002WR001879}

\bibitem[{Stepinski and Vilalta(2005)}]{Stepinski2005}
Stepinski, T., Vilalta, R., 2005. Digital topography models for martian
  surfaces. Geoscience and Remote Sensing Letters, IEEE 2~(3), 260--264.
 doi: \url{http://dx.doi.org/10.1109/LGRS.2005.848509}

\bibitem[{Tarboton(1997)}]{tarboton1997}
Tarboton, D., 1997. A new method for the determination of flow directions and
  upslope areas in grid digital elevation models. Water resources research 33,
  309--319.
 doi: \url{http://dx.doi.org/10.1029/96WR03137}

\bibitem[{Tarboton and Ames(2001)}]{Tarboton2001}
Tarboton, D., Ames, D., 2001. Advances in the mapping of flow networks from
  digital elevation data. In: World Water and Environmental Resources Congress.
  Am. Soc Civil Engrs USA, pp. 20--24.
 doi: \url{http://dx.doi.org/10.1061/40569(2001)166}

\bibitem[{Toma et~al.(2001)Toma, Wickremesinghe, Arge, Chase, Vitter, Halpin,
  and Urban}]{Toma2001}
Toma, L., Wickremesinghe, R., Arge, L., Chase, J.~S., Vitter, J.~S., Halpin,
  P.~N., Urban, D., 2001. Flow computation on massive grids. In: Proceedings of
  the 9th ACM international symposium on Advances in geographic information
  systems. GIS '01. ACM, New York, NY, USA, pp. 82--87.
 doi: \url{http://dx.doi.org/10.1145/512161.512180}

\bibitem[{Tribe(1992)}]{Tribe1992}
Tribe, A., 1992. Automated recognition of valley lines and drainage networks
  from grid digital elevation models: a review and a new method. Journal of
  Hydrology 139~(1-4), 263--293.
 doi: \url{http://dx.doi.org/10.1016/0022-1694(92)90206-B}

\bibitem[{Turcotte et~al.(2001)Turcotte, Fortin, Rousseau, Massicotte, and
  Villeneuve}]{Turcotte2001}
Turcotte, R., Fortin, J., Rousseau, A., Massicotte, S., Villeneuve, J., 2001.
  Determination of the drainage structure of a watershed using a digital
  elevation model and a digital river and lake network. Journal of Hydrology
  240~(3-4), 225--242.
 doi: \url{http://dx.doi.org/10.1016/S0022-1694(00)00342-5}

\bibitem[{Vogt et~al.(2003)Vogt, Colombo, and Bertolo}]{Vogt2003}
Vogt, J., Colombo, R., Bertolo, F., 2003. Deriving drainage networks and
  catchment boundaries: a new methodology combining digital elevation data and
  environmental characteristics. Geomorphology 53~(3-4), 281--298.
 doi: \url{http://dx.doi.org/10.1016/S0169-555X(02)00319-7}

\bibitem[{Wallis et~al.(2009)Wallis, Watson, Tarboton, and
  Wallace}]{Wallis2009}
Wallis, C., Watson, D., Tarboton, D., Wallace, R., 2009. Parallel
  flow-direction and contributing area calculation for hydrology analysis in
  digital elevation models. In: International Conference on Parallel and
  Distributed Processing Techniques and Applications.

\bibitem[{Wang and Liu(2006)}]{wang2006}
Wang, L., Liu, H., 2006. An efficient method for identifying and filling
  surface depressions in digital elevation models for hydrologic analysis and
  modelling. International Journal of Geographical Information Science 20~(2),
  193--213.
 doi: \url{http://dx.doi.org/10.1080/13658810500433453}

\bibitem[{White et~al.(2004)White, Kumar, Saco, Rhoads, and Yen}]{White2004}
White, A., Kumar, P., Saco, P., Rhoads, B., Yen, B., 2004. Hydrodynamic and
  geomorphologic dispersion: scale effects in the illinois river basin. Journal
  of hydrology 288~(3-4), 237--257.
 doi: \url{http://dx.doi.org/10.1016/j.jhydrol.2003.10.019}

\bibitem[{Zhang and Huang(2009)}]{Zhang2009}
Zhang, H., Huang, G., 2009. Building channel networks for flat regions in
  digital elevation models. Hydrological Processes 23~(20), 2879--2887.
 doi: \url{http://dx.doi.org/10.1002/hyp.7378}

\end{thebibliography}
}
\newpage

\appendix
\section{Pseudocode}

\begin{algorithm}
\caption{{\sc ResolveFlats}: The main body of the algorithm, as described in~\textsection\ref{sec:overview}. \textbf{Upon entry,} \textbf{(1)}~\textit{DEM} contains the elevations of every cell or a value \textsc{NoData} for cells not part of the DEM. \textbf{(2)}~\textit{FlowDirs} contains the flow direction of every cell; cells without a local gradient are marked \textsc{NoFlow}. Algorithm~\ref{alg:d8flowdirs} provides an example of how this might be done. \textbf{At exit,} \textbf{(1)}~\textit{FlatMask} has a value greater than or equal to zero for each cell, indicating its number of increments. These can be used be used in conjunction with \textit{Labels} to determine flow directions without altering the DEM, as exemplified by Algorithm~\ref{alg:d8maskedflowdirs}, or to alter the DEM in subtle ways to direct flow, as exemplified by Algorithm~\ref{alg:alterdem}. \textbf{(2)}~\textit{Labels} has values greater than or equal to~1 for each cell which is in a flat. Each flats' cells bear a label unique to that flat.}
\label{alg:resolveflats}
\begin{algorithmic}[1]
  \Procedure{ResolveFlats}{\textit{DEM}, \textit{FlowDirs}}
  \Require The dimensions of $\textit{DEM}$ must equal those of $\textit{FlowDirs}$
	\State Let \textit{FlatMask} and \textit{Labels} have the same dimensions as \textit{DEM} and be initialized to 0
	\State Let \textit{LowEdges} and \textit{HighEdges} be two empty FIFO queues
	\State
	\State \Call{FlatEdges}{\textit{HighEdges},\textit{LowEdges}} \Comment Algorithm~\ref{alg:findedges}
	\If{\textit{LowEdges} is empty}
		\If{\textit{HighEdges} is not empty}
			\State There were undrainable flats
		\Else
			\State There were no flats
		\EndIf
		\State \Return
	\EndIf
	\State
	\State $\textit{Label}\gets1$
	\ForAll{$c\in\textit{LowEdges}$}
		\If{$c$ is not labeled}
			\State \Call{LabelFlats}{$c,\textit{Label}$} \Comment Algorithm~\ref{alg:label}
			\State $\textit{Label}\gets\textit{Label}+1$
		\EndIf
	\EndFor
	\State
	\ForAll{$c\in\textit{HighEdges}$}
		\LineIf{$c$ is not labeled}{Remove $c$}
	\EndFor
	\If{Any cell was removed from \textit{HighEdges}}
		\State Not all flats have outlets
	\EndIf
	\State
	\State Let \textit{FlatHeight} be an array with length equal to the value of \textit{Label}
	\State \Call{AwayFromHigher}{\textit{HighEdges}, \textit{FlatHeight}} \Comment Algorithm~\ref{alg:higher_grad}
	\State \Call{TowardsLower}{\textit{LowEdges}, \textit{FlatHeight}} \Comment Algorithm~\ref{alg:lower_grad}
	\State \Return \textit{FlatMask},\textit{Labels}
  \EndProcedure
\end{algorithmic}
\end{algorithm}

\begin{algorithm}
\caption{{\sc D8FlowDirs}: This is an example of how flow directions could be assigned to cells, as described in~\textsection\ref{sec:flatid}. \textbf{Upon entry,} \textbf{(1)}~\textit{DEM} contains the elevations of each cell or a value \textsc{NoData} for cells not part of the DEM. \textbf{At exit,} \textit{FlowDirs} contains a flow direction for each cell which has a local gradient; those cells without a local gradient are marked \textsc{NoFlow}.}
\label{alg:d8flowdirs}
\begin{algorithmic}[1]
	\Procedure{D8FlowDirs}{\textit{DEM}, \textit{FlowDirs}}
  \Require \textit{FlowDirs} has the same dimensions as \textit{DEM}
	\ForAll{$c$ in \textit{DEM}}
		\If{$\textit{DEM}(c)=\textsc{NoData}$}
      \State $FlowDirs(c)\gets\textsc{NoData}$
      \State \Continue
    \EndIf
    \State $e_\textrm{min}\gets\textit{DEM}(c)$
    \State $n_\textrm{min}\gets\textsc{NoFlow}$
		\ForAll{neighbors $n$ of $c$}
			\LineIf{$n\notin\textit{DEM}$}{\Continue}
      \If{$\textit{DEM}(n)<e_\textrm{min}$}
				\State $e_\textrm{min}\gets\textit{DEM}(n)$
        \State $n_\textrm{min}\gets$ direction to $n$
			\EndIf
		\EndFor
    \State $\textit{FlowDirs}(c)\gets n_\textrm{min}$
	\EndFor
	\EndProcedure
\end{algorithmic}
\end{algorithm}

\begin{algorithm}
\caption{{\sc FlatEdges}: This function locates flat cells which border on higher and lower terrain and places them into queues for further processing, as described in~\textsection\ref{sec:flatid}. \textbf{Upon entry,} \textbf{(1)}~\textit{DEM} contains the elevations of every cell or a value \textsc{NoData} for cells not part of the DEM. \textbf{(2)}~Any cell without a local gradient is marked \textsc{NoFlow} in \textit{FlowDirs}. \textbf{At exit,} \textbf{(1)}~\textit{HighEdges} contains all the high edge cells (those flat cells adjacent to higher terrain) of the DEM, in no particular order. \textbf{(2)}~\textit{LowEdges} contains all the low edge cells of the DEM, in no particular order.} 
\label{alg:findedges}
\begin{algorithmic}[1]
	\Procedure{FlatEdges}{\textit{HighEdges}, \textit{LowEdges}}
	\Require \textit{DEM},\textit{FlowDirs}
	\ForAll{$c$ in \textit{Flowdirs}}
		\LineIf{$c=\textsc{NoData}$}{\Continue}
		\ForAll{neighbors $n$ of $c$}
			\LineIf{$n\notin\textit{DEM}$}{\Continue}
			\LineIf{$n=\textsc{NoData}$}{\Continue}
			\If{$c\ne\textsc{NoFlow}$ \aand $n=\textsc{NoFlow}$ \aand $DEM(c)=DEM(n)$}
				\State Push $c$ onto $\textit{LowEdges}$
				\Break
			\ElsIf{$c=\textsc{NoFlow}$ \aand $DEM(c)<DEM(n)$}
				\State Push $c$ onto $\textit{HighEdges}$
				\Break
			\EndIf
		\EndFor
	\EndFor
	\EndProcedure
\end{algorithmic}
\end{algorithm}

\begin{algorithm}
\caption{{\sc LabelFlats}: This flood-fill function gives all the cells of a flat a common label, as described by~\textsection\ref{sec:flatid}. \textbf{Upon entry,} \textbf{(1)}~\textit{DEM} contains the elevations of every cell or a value \textsc{NoData} for cells not part of the DEM. \textbf{(2)}~\textit{Labels} has the same dimensions as \textit{DEM}. \textbf{(3)}~$c$ belongs to the flat which is to be labeled. \textbf{(4)}~$L$ is a unique label which has not been previously applied to a flat. \textbf{(5)}~\textit{Labels} has been initialized to zero prior to the first call to this function. \textbf{(6)}~\textit{Labels} has values greater than or equal to~1 for each processed cell which is in a flat. Each flats' cells bear a label unique to that flat. \textbf{At exit,} \textbf{(1)}~$c$ and every cell reachable from $c$ by passing over only cells of the same elevation as $c$ (all the cells in the flat to which $c$ belongs) is marked as $L$ in \textit{Labels}. \textbf{(2)}~\textit{Labels} has been updated to reflect the new labels which have been applied.} 
\label{alg:label}
\begin{algorithmic}[1]
\Procedure{LabelFlats}{cell $c$, label $L$}
\Require $\textit{DEM}, \textit{Labels}$
	\State Let \textit{ToFill} be an empty FIFO queue
	\State Push $c$ onto $\textit{ToFill}$
	\State $E\gets\textit{DEM}(c)$
	\While{$\textit{ToFill}$ is not empty}
		\State $c\gets\textsc{pop}(\textit{ToFill}$)
		\LineIf{$c\notin\textit{DEM}$}{\Continue}
		\LineIf{$\textit{DEM}(c)\ne E$}{\Continue}\Comment $c$ is not a part of the flat
		\LineIf{$\textit{Labels}(c)\ne0$}{\Continue}\Comment $c$ is already labeled
		\State $\textit{Labels}(c)\gets L$
		\State Push all neighbors of $c$ onto $\textit{ToFill}$
	\EndWhile
\EndProcedure
\end{algorithmic}
\end{algorithm}

\begin{algorithm}
\caption{{\sc AwayFromHigher}: This procedure builds a gradient away from higher terrain, as described in~\textsection\ref{sec:higher_grad} and Fig.~\ref{fig:higher_grad}. \textbf{Upon entry,} \textbf{(1)}~Every cell in \textit{Labels} is marked either 0, indicating that the cell is not part of a flat, or a number greater than zero which identifies the flat to which the cell belongs. \textbf{(2)}~Any cell without a local gradient is marked \textsc{NoFlow} in \textit{FlowDirs}. \textbf{(3)}~Every cell in \textit{FlatMask} is initialized to 0. \textbf{(4)}~\textit{HighEdges} contains, in no particular order, all the high edge cells of the DEM which are part of drainable flats. \textbf{At exit,} \textbf{(1)}~\textit{FlatHeight} has an entry for each label value of \textit{Labels} indicating the maximal number of increments to be applied to the flat identified by that label. \textbf{(2)}~\textit{FlatMask} contains the number of increments to be applied to each cell to form a gradient away from higher terrain; cells not in a flat have a value of 0.}
\label{alg:higher_grad}
\begin{algorithmic}[1]
	\Procedure{AwayFromHigher}{\textit{HighEdges}, \textit{FlatHeight}}
	\Require \textit{Labels}, \textit{FlatMask}, \textit{FlowDirs}
  \State Let \textit{FlatMask} have the same dimensions as \textit{Labels}
  \State Let \textit{FlatHeight} have length equal to \textsc{Max}(\textit{Labels})
	\State $\textit{loops}\gets1$
	\State Push {\sc Marker} onto $HighEdges$
	\While{$\textsc{length}(\textit{HighEdges})>1$}
		\State $c\gets\textsc{pop}(\textit{HighEdges})$
		\If{$c=\textsc{Marker}$}
			\State $loops\gets loops+1$
			\State Push {\sc Marker} onto $HighEdges$
			\State \Continue
		\EndIf
		\LineIf{$\textit{FlatMask}(c)>0$}{\Continue}
		\State $\textit{FlatMask}(c)\gets \textit{loops}$
		\State $\textit{FlatHeight}(\textit{Labels}(c))\gets \textit{loops}$
		\ForAll{neighbors $n$ of $c$}
			\If{$n\in\textit{Labels}$ \aand $\textit{Labels}(n)=\textit{Labels}(c)$ \aand $\textit{FlowDirs}(n)=\textsc{NoFlow}$}
				\State Push $n$ onto \textit{HighEdges}
			\EndIf
		\EndFor
	\EndWhile
	\EndProcedure
\end{algorithmic}
\end{algorithm}

\begin{algorithm}
\caption{{\sc TowardsLower}: This procedure builds a gradient towards lower terrain and combines it with the gradient away from higher terrain, as described in~\textsection\ref{sec:lower_grad} and Fig.~\ref{fig:lower_grad}. \textbf{Upon entry,} \textbf{(1)}~Every cell in \textit{Labels} is marked either 0, indicating that the cell is not part of a flat, or a number greater than zero which identifies the flat to which the cell belongs. \textbf{(2)}~Any cell without a local gradient is marked \textsc{NoFlow} in \textit{FlowDirs}. \textbf{(3)}~Every cell in \textit{FlatMask} has either a value of 0, indicating that the cell is not part of a flat, or a value greater than zero indicating the number of increments which must be added to it to form a gradient away from higher terrain. \textbf{(4)}~\textit{FlatHeight} has an entry for each label value of \textit{Labels} indicating the maximal number of increments to be applied to the flat identified by that label in order to form the gradient away from higher terrain. \textbf{(5)}~\textit{LowEdges} contains, in no particular order, all the low edge cells of the DEM. \textbf{At exit,} \textbf{(1)}~\textit{FlatMask} contains the number of increments to be applied to each cell to form a superposition of the gradient away from higher terrain with the gradient towards lower terrain; cells not in a flat have a value of 0.}
\label{alg:lower_grad}
\begin{algorithmic}[1]
	\Procedure{TowardsLower}{\textit{LowEdges}, \textit{FlatHeight}}
	\Require \textit{Labels}, \textit{FlatMask}, \textit{FlowDirs}
	\State Make all entries in \textit{FlatMask} negative
	\State $\textit{loops}\gets1$
	\State Push {\sc Marker} onto $LowEdges$
	\While{$\textsc{length}(\textit{LowEdges})>1$}
		\State $c\gets\textsc{pop}(\textit{LowEdges})$
		\If{$c=\textsc{Marker}$}
			\State $loops\gets loops+1$
			\State Push {\sc Marker} onto $LowEdges$
			\State \Continue
		\EndIf
		\LineIf{$\textit{FlatMask}(c)>0$}{\Continue}
		\If{$\textit{FlatMask}(c)<0$}
			\State $\textit{FlatMask}(c)\gets\textit{FlatHeight}(\textit{Labels}(c))+FlatMask(c)+2\cdot\textit{loops}$
		\Else
			\State $\textit{FlatMask}(c)\gets2\cdot\textit{loops}$
		\EndIf

		\ForAll{neighbors $n$ of $c$}
			\If{$n\in\textit{Labels}$ \aand $\textit{Labels}(n)=\textit{Labels}(c)$ \aand $\textit{FlowDirs}(n)=\textsc{NoFlow}$}
				\State Push $n$ onto \textit{LowEdges}
			\EndIf
		\EndFor
	\EndWhile
	\EndProcedure
\end{algorithmic}
\end{algorithm}

\begin{algorithm}
\caption{{\sc D8MaskedFlowDirs}: This is an example of how the \textit{FlatMask} and \textit{Labels} developed previously may be used to determine flow directions across flats, as described in~\textsection\ref{sec:flowdir}. This function will need to be adjusted to fit with the flow metric being used. \textbf{Upon entry,} \textbf{(1)}~\textit{FlatMask} contains the number of increments to be applied to each cell to form a gradient which will drain the flat it is a part of. \textbf{(2)}~Any cell without a local gradient has a value of \textsc{NoFlow} in \textit{FlowDirs}; all other cells have defined flow directions. \textbf{(3)}~If a cell is part of a flat, it has a value greater than zero in \textit{Labels} indicating which flat it is a member of; otherwise, it has a value of 0. \textbf{At exit,} every cell whose flow direction could be resolved by this algorithm (all drainable flats) has a defined flow direction in \textit{FlowDirs}. Any cells which could not be resolved (non-drainable flats) are still marked \textsc{NoFlow}.}
\label{alg:d8maskedflowdirs}
\begin{algorithmic}[1]
	\Procedure{D8MaskedFlowDirs}{\textit{FlatMask}, \textit{FlowDirs}, \textit{Labels}}
	\ForAll{$c$ in \textit{FlowDirs}}
		\LineIf{$\textit{FlowDirs}(c)=\textsc{NoData}$}{\Continue}
		\LineIf{$\textit{FlowDirs}(c)\ne\textsc{NoFlow}$}{\Continue}
    \State $e_\textrm{min}\gets\textit{FlatMask}(c)$
    \State $n_\textrm{min}\gets\textsc{NoFlow}$
		\ForAll{neighbors $n$ of $c$}
			\LineIf{$n\notin\textit{FlowDirs}$}{\Continue}
			\LineIf{$\textit{Labels}(n)\ne\textit{Labels}(c)$}{\Continue}
      \If{$\textit{FlatMask}(n)<e_\textrm{min}$}
				\State $e_\textrm{min}\gets\textit{FlatMask}(n)$
        \State $n_\textrm{min}\gets$ direction to $n$
			\EndIf
		\EndFor
    \State $\textit{FlowDirs}(c)\gets n_\textrm{min}$
	\EndFor
	\EndProcedure
\end{algorithmic}
\end{algorithm}

\begin{algorithm}
\caption{{\sc AlterDEM}: This is an example of how the \textit{FlatMask} and \textit{Labels} developed previously may be used to increase the elevation of cells in the original DEM to enforce drainage directions, as described in~\textsection\ref{sec:flowdir}. The \textsc{NextAfter} function is defined by the \texttt{C99} and \texttt{POSIX} standards. A safety check is used to ensure that this doesn't change the original hydrology of the DEM. \textbf{Upon entry,} \textbf{(1)}~\textit{FlatMask} contains the number of increments to be applied to each cell to form a gradient which will drain the flat. \textbf{(2)}~Any cell without a local gradient has a value of \textsc{NoFlow} in \textit{FlowDirs}; all other cells have defined flow directions. \textbf{(3)}~If a cell is part of a flat, it has a value greater than zero in \textit{Labels} indicating which flat it is a member of; otherwise, it has a value of 0. \textbf{At exit,} \textit{DEM} has been altered so that the elevation of every cell which was part of a drainable flat is adjusted such that it is guaranteed to drain.}
\label{alg:alterdem}
\begin{algorithmic}[1]
	\Procedure{AlterDEM}{\textit{DEM}, \textit{FlatMask}, \textit{Labels}}
	\ForAll{$c$ in \textit{DEM}}
		\LineIf{$\textit{DEM}(c)=\textsc{NoData}$}{\Continue}
		\LineIf{$\textit{Label}(c)=0$}{\Continue}\Comment Cell is not part of a flat
		\LineIf{$\textit{FlatMask}(c)=0$}{\Continue}
		\ForAll{neighbors $n$ of $c$}\Comment Note whether $c$ is already higher than $n$
      \State $\textit{Higher}(n)\gets\left(\textit{DEM}(c)>\textit{DEM}(n)\right)$
    \EndFor
    \For{$j=0 \to \textit{FlatMask}(c)$}\Comment Alter the DEM
      \State $\textit{DEM}(c)\gets$\Call{NextAfter}{$\textit{DEM}(c),\infty$}
    \EndFor
		\ForAll{neighbors $n$ of $c$}\Comment Check if alteration was significant
      \LineIf{$\textit{Labels}(n)=\textit{Labels}(c)$}{\Continue}
      \LineIf{$\textit{DEM}(c)<\textit{DEM}(n)$}{\Continue}
      \If{\textbf{not} $\textit{Higher}(n)$}
        \State $c$ was lower than $n$ before, but it isn't now.
        \State The alteration was significant.
        \State This is a problem.
      \EndIf
    \EndFor
  \EndFor
	\EndProcedure
\end{algorithmic}
\end{algorithm}

\end{document}